\documentclass[12pt]{article}

\usepackage{times}

\DeclareFontFamily{OT1}{times}{}
\DeclareFontShape{OT1}{times}{m}{n }{ <-> ptmr }{}
\DeclareFontShape {OT1}{times}{bx}{n }{ <-> ptmb }{}
\DeclareFontShape {OT1}{times}{m }{it}{ <-> ptmri}{}
\DeclareFontShape {OT1}{times}{bx}{it}{ <-> ptmbi}{}
\usepackage{amsmath}
\usepackage{amsfonts}
\usepackage{amssymb}
\usepackage{latexsym}
\usepackage{graphics}

\begin{document}

\newtheorem{theo}{Theorem}[section]
\newtheorem{defi}[theo]{Definition}
\newtheorem{prop}[theo]{Proposition}
\newtheorem{corr}[theo]{Corollary}
\newtheorem{lemm}[theo]{Lemma}
\newtheorem{exam}[theo]{Example}

\newcommand{\CD}[2]{\ensuremath{\raisebox{-.6ex}{\scriptsize{$0$}}\:\!
        \mathbf{C}_{t}^{#1}\left[#2\right]}}
\newcommand{\FFD}[4]{\ensuremath{\raisebox{-.6ex}{\scriptsize{$#1$}}\!
        \raisebox{1ex}{\scriptsize{$t$}}\:\!\mathbf{F}_{#2}^{#3}[#4]}}
\newcommand{\myH}[1]{\ensuremath{H^{#1}([0,T])}}
\newcommand{\K}[3]{\ensuremath{K_{#1}^{(#2)}(#3)}}
\newcommand{\Kstar}[3]{\ensuremath{K_{#1}^{(#2)} \star q^{#3}}}
\newcommand{\myL}{\ensuremath{L^{2}([0,T])}}
\newcommand{\LFD}[3]{\ensuremath{\raisebox{-.6ex}{\scriptsize{$#1$}}\:\!
        \mathbf{D}_{t}^{#2}\left[#3\right]}}
\newcommand{\LFDt}[3]{\ensuremath{\raisebox{-.6ex}{\scriptsize{$#1$}}\:\!
        \mathbf{D}_{\tau}^{#2}\left[#3\right]}}
\newcommand{\LFDm}[3]{\ensuremath{\raisebox{-.6ex}{\scriptsize{$#1$}}\:\!
        \mathbf{D}_{t^{-}}^{#2}\left[#3\right]}}
\newcommand{\LFI}[3]{\ensuremath{\raisebox{-.6ex}{\scriptsize{$#1$}}\:\!
        \mathbf{I}_{t}^{#2}[#3]}}
\newcommand{\mysum}{\ensuremath{\sum_{n=1}^{\infty}}}
\newcommand{\PHIm}[2]{\ensuremath{\Phi_{#1}^{-}(#2)}}
\newcommand{\PHIp}[2]{\ensuremath{\Phi_{#1}^{+}(#2)}}
\newcommand{\PSI}[2]{\ensuremath{\Psi_{#1}(#2)}}
\newcommand{\lfeta}[2]{\ensuremath{\raisebox{-.6ex}{\scriptsize{$#1$}}\:\!\eta_{t}^{#2}}}
\newcommand{\rfeta}[2]{\ensuremath{\raisebox{-.6ex}{\scriptsize{$t$}}\:\!\eta_{#1}^{#2}}}
\newcommand{\lfq}[2]{\ensuremath{\raisebox{-.6ex}{\scriptsize{$#1$}}\:\!q_{t}^{#2}}}
\newcommand{\RFD}[3]{\ensuremath{\raisebox{-.6ex}{\scriptsize{$t$}}\:\!
        \mathbf{D}_{#1}^{#2}\left[#3\right]}}
\newcommand{\RFDp}[3]{\ensuremath{\raisebox{-.6ex}{\scriptsize{$t^{+}$}}\:\!
        \mathbf{D}_{#1}^{#2}\left[#3\right]}}
\newcommand{\RFI}[3]{\ensuremath{\raisebox{-.6ex}{\scriptsize{$t$}}\:\!
        \mathbf{I}_{$#1$}^{#2}[#3]}}
\newcommand{\RLD}[2]{\ensuremath{\raisebox{-.6ex}{\scriptsize{$0$}}\:\!
        \mathbf{R}_{t}^{#1}\left[#2\right]}}
\newcommand{\Rk}[2]{\ensuremath{R^{#1}_{#2}(\tau)}}
\newcommand{\rk}[2]{\ensuremath{r^{#1}_{#2}(\tau)}}
\newcommand{\SIP}[3]{\ensuremath{\langle q_{#1},q_{#2}\rangle_{H^{#3}}}}
\newcommand{\Test}{\ensuremath{C^{\infty}([0,T])}}


\title{Nonconservative Lagrangian mechanics II: purely causal equations of motion}

\author{David W. Dreisigmeyer\thanks{email:dreisigm@math.colostate.edu}\\
        Department of Mathematics\\
        Colorado State University, Fort Collins, CO 80523
        \and
        Peter M. Young\thanks{email:pmy@engr.colostate.edu}\\
        Department of Electrical and Computer Engineering\\
        Colorado State University, Fort Collins, CO 80523}

\maketitle
\begin{abstract}
This work builds on the Volterra series formalism presented in [D.
W. Dreisigmeyer and P. M. Young, J. Phys. A \textbf{36}, 8297,
(2003)] to model nonconservative systems.  Here we treat
Lagrangians and actions as `time dependent' Volterra series.  We
present a new family of kernels to be used in these Volterra
series that allow us to derive a single retarded equation of
motion using a variational principle.\\

\noindent PACS: 45.20.-d, 02.30.-f
\end{abstract}

\section{Introduction}
\label{SEC-Introduction}

The central question addressed in this paper is: `How can one have
a single retarded equation of motion arise from the use of a
variational principle?'  Having a dissipative equation of motion
arise from the use of a variational principle has a rather long
history in mechanics.  One of the central results in this area is
Bauer's 1931 corollary \cite{Baue31}:
\begin{corr}\label{theo1}
The equations of motion of a dissipative linear dynamical system
with constant coefficients are not given by a variational
principle.
\end{corr}
Bauer's proof of this corollary relies on two implicit
assumptions.  First, there are no extra equations of motion that
arise.  Also, only integer ordered derivative operators are used
in the action and Lagrangian.  It is the latter assumption that we
will eventually take advantage of to avoid Bauer's corollary.

There has been some effort devoted to bypassing Bauer's result.
One of the earliest attempts was by Bateman \cite{Bate31}.  What
he did was allow a dual equation of motion to arise.  Let us
illustrate this via the dissipative harmonic oscillator.  If we
start with the Lagrangian
\begin{eqnarray}\label{BateLag}
L & = & m \dot{x} \dot{y} + \frac{C}{2} (x\dot{y} - \dot{x} y) - m
    \omega^{2} x y      ,
\end{eqnarray}
where C is a constant, we would have the following equations of
motion
\begin{eqnarray}
m \ddot{x} + C \dot{x} + m \omega^{2}x & = & 0 \label{xeom}  \\
m \ddot{y} - C \dot{y} + m \omega^{2}y & = & 0 \label{yeom}\ .
\end{eqnarray}
Equation (\ref{xeom}) is what we want for our harmonic oscillator
(i.e., it is retarded or causal). Equation (\ref{yeom}) is a time
reversed version of (\ref{xeom}). Bateman's method is not
particularly general. Also, the appearance of an advanced (time
reversed or anti-causal) equation can be considered a drawback to
this procedure.  Our universe appears to be causal so there does
not seem to be any compelling reason why anti-causal effects
should arise from any correct action.

A rather novel attempt to get around Bauer's corollary was
explored by Riewe in a series of papers \cite{Riew96,Riew97}.
Riewe tried using fractional derivatives in the action to have
nonconservative equations arise.  In particular, Riewe did not use
Bateman's method of having an anti-causal dual equation. However,
Riewe's equations of motion are acausal.  In order to circumvent
this, it was suggested that all anti-causal operators be replaced
with their causal counterparts.  (Riewe's method is rather
involved so we do not provide any illustrative examples.) This
procedure does allow for much more general equations than
Bateman's method.  However, it must be remembered that the actual
equations derived are acausal and the procedure to change these
into causal equations is rather ad hoc.

In examining Riewe's approach, Dreisigmeyer and Young showed that
his procedure of replacing anti-causal with causal operations may
not be a wise idea \cite{DY03}.  While still employing fractional
derivatives, Dreisigmeyer and Young instead allowed for a causal
equation and an anti-causal dual equation to arise.  Thus, their
procedure can be considered a generalization of Bateman's.  The
appearance of an anti-causal equation is still very troublesome.
Since both \cite{Riew96,Riew97} and \cite{DY03} deal with
fractional derivatives, we will later refer to these techniques as
\textit{fractional mechanics}.

There are some other approaches to our basic problem.  Tonti
\cite{Tont71} and Arthurs and Jones \cite{ArJo76} did develop a
procedure based on the convolution product that is useful for the
harmonic oscillator. However, this does not seem to generalize to
higher ordered potentials.  So we need to question if this is the
correct procedure to follow.  Caldeira and Leggett \cite{CaLe83}
modelled nonconservative systems by coupling them to an
environment.  The environment is modelled as a collection of
harmonic oscillators which results in the Lagrangian
\begin{eqnarray}\label{CLmod}
L & = & \frac{m}{2}\ \dot{q}^{2} - V(q) + \sum_{n=1}^{\infty}
        \left\{ \frac{m_{n}}{2}\ \dot{q}_{n}^{2} - \frac{m_{n}
        \omega_{n}^{2}}{2}\ (q_{n}-q)^{2} \right\}
\end{eqnarray}
where $q$ is the system's coordinate and the $q_{n}$ are the
environment's coordinates.  Here the combined system and
environment is conservative while the system alone may be
dissipative.  This procedure allows for the introduction of very
general dissipation terms into the system's equation of motion.
For example, fractional derivatives can be included.  However, the
microscopic modelling of the environment makes (\ref{CLmod})
rather complicated.  In particular, we would generally have a
countably infinite system of equations resulting from
(\ref{CLmod}).  Also, the question of causality is `put off' to
the end. That is, we enforce causality in $q$'s equation of motion
by only using the causal solutions of the $q_{n}$'s equations of
motion.  In other words, we assume the environment acts causally
only when we are solving it's equations of motion.  While this is
not particularly disturbing, it does mean that causality is not
necessarily enforced in the action.

In \cite{DY03} the idea of treating the actions and Lagrangians as
Volterra series was introduced.  (Volterra series are the
generalization to functionals of the power series concept.)  This
is a rather powerful framework. It allows us to build up
Lagrangians (and hence actions) in a rather systematic way by
tailoring the kernels in the Volterra series to our requirements.
As originally stated, we desire to derive a single causal equation
of motion for a system.  We  will accomplish this by choosing the
correct family of kernels for our Volterra series.  Our paper is
organized as follows.  We review the distributional approach to
fractional differentiation in section \ref{SEC-FracDer}.  In
section \ref{SEC-Volterra} we introduce Volterra series. Our
variational principle is developed in section \ref{SEC-Causal}.  A
discussion of our results and possible future research follows in
section \ref{SEC-Discussion}.

\section{Fractional derivatives}\label{SEC-FracDer}

We will now briefly review fractional derivatives using the
distributional approach. (A fuller discussion of this material can
be found in \cite{DY03,GeSh64,Podl99}.) First, define the
generalized functions
\begin{eqnarray}\label{PHIpdef}
\PHIp{\alpha}{t} & = & \left\{ \begin{array}{cc}
                \frac{1}{\Gamma(\alpha)}\ t^{\alpha-1} & t>0 \\
                0 & t \leq 0
                \end{array} \right.
\end{eqnarray}
and
\begin{eqnarray}\label{PHImdef}
\PHIm{\alpha}{t} & = & \left\{ \begin{array}{cc}
                \frac{1}{\Gamma(\alpha)}\ |t|^{\alpha-1} & t<0 \\
                0 & t\geq 0
                \end{array} \right.\
\end{eqnarray}
where $\Gamma(\alpha)$ is the gamma function.  These distributions
will allow us to define two different types of fractional
derivatives.  One of these will be causal while the other will be
anti-causal.  The anti-causal derivatives need to be avoided in
our equations of motion.

Left fractional derivatives (LFDs) of order $\alpha$ of a function
$q(t)$ are defined by
\begin{eqnarray}\label{LFDdef}
\LFD{a}{\alpha}{q} & := & \PHIp{-\alpha}{t} \ast q(t)    \\
    & = & \frac{1}{\Gamma(-\alpha)}\ \int_{a}^{t} q(\tau)
    (t-\tau)^{-(\alpha + 1)}\ d\tau   \nonumber
\end{eqnarray}
where we set $q(t) \equiv 0$ for $t<a$.  For $\alpha = n$, $n$ an
integer, (\ref{LFDdef}) becomes
\begin{eqnarray}\label{LFDint}
\LFD{a}{n}{q} & = & \mathcal{D}^{n} q
\end{eqnarray}
where $\mathcal{D}$ is the generalized derivative.  The LFDs are
causal operations.  Hence, we will only want LFDs to be in our
equations of motion.

Right fractional derivatives (RFDs) of order $\alpha$ are given by
\begin{eqnarray}\label{RFDdef}
\RFD{b}{\alpha}{q} & := & \PHIm{-\alpha}{t} \ast q(t)    \\
    & = & \frac{1}{\Gamma(-\alpha)} \int_{t}^{b} q(\tau) (\tau -
    t)^{-(\alpha + 1)}\ d\tau     \nonumber
\end{eqnarray}
where now $q(t) \equiv 0$ for $t>b$.  Notice that instead of
(\ref{LFDint}), we have
\begin{eqnarray}\label{RFDint}
\RFD{b}{n}{q} & = & (-1)^{n} \mathcal{D}^{n} q\ .
\end{eqnarray}
The RFDs are anti-causal operations.  So, we do not want the RFDs
to appear in our equations of motion.

The basic problem for fractional mechanics is to avoid the
appearance of RFDs in the equations of motion.  Neither Riewe
\cite{Riew96,Riew97} nor Dreisigmeyer and Young \cite{DY03} were
able to remove RFDs completely.  Riewe suggested replacing RFDs
with LFDs in the resulting equations of motion.  Dreisigmeyer and
Young suggested letting the LFDs appear in one equation while the
RFDs appeared in a dual equation.  Neither of these methods is
entirely satisfactory.  We will shortly present a formalism that
completely avoids RFDs.  Before that, let us look at Volterra
series.

\section{Volterra Series}\label{SEC-Volterra}

Now we are going to develop some of the Volterra series concepts
that we will need.  Our treatment of Volterra series here is
somewhat different than that presented in \cite{DY03}. Here we
treat Volterra series as expansions of (generalized) functions
that depend on time.  This is the viewpoint adopted in the
nonlinear systems theory that is somewhat popular in electrical
engineering (see, e.g., \cite{Rugh02}).  In \cite{DY03} the
Volterra series were treated as expansions of functionals only
(i.e., no time dependence). This change in viewpoint will make all
the difference in achieving our goal of a single causal equation
of motion.

For completeness let us first review the Volterra series treatment
in \cite{DY03}, see also \cite{Stev95}.  (Please note, we will not
review the fractional mechanics formalism developed by
Dreisigmeyer and Young in this paper since this will be largely
irrelevant for our current work and also take us too far afield.)
In \cite{DY03} only functionals were treated as (time independent)
Volterra series. For some functional $\mathcal{V}[q]$ define the
kernels
\begin{eqnarray}\label{Ks}
\K{n}{s}{\tau_{1},\ldots,\tau_{n}} & := & \frac{\delta^{n}
        \mathcal{V}[q]}
        {\delta q(\tau_{1}) \cdots \delta q(\tau_{n})}  .
\end{eqnarray}
Notice that the $\K{n}{s}{\cdot}$'s are symmetric under an
interchange of the $\tau_{i}$'s.  For example,
$\K{2}{s}{\tau_{1},\tau_{2}} = \K{2}{s}{\tau_{2},\tau_{1}}$.  Now
introduce the convenient notation
\begin{eqnarray}\label{Knstar}
\Kstar{n}{s}{n} & := & \int_{\tau_{1}} \cdots \int_{\tau_{n}}
        \K{n}{s}{\tau_{1},\ldots,\tau_{n}} q(\tau_{n}) \cdots
        q(\tau_{1})\ d\tau_{n} \cdots d\tau_{1}  .
\end{eqnarray}
Then we can represent the functional $\mathcal{V}[q]$ as the
Volterra series
\begin{eqnarray}\label{VolterraSeries}
\mathcal{V}[q] & = & \mysum \frac{1}{n!} \Kstar{n}{s}{n}    .
\end{eqnarray}
The $\mathcal{V}[q]$ were taken as the actions in \cite{DY03}.
(This is why we can take $K_{0}^{(s)} = \mathcal{V}[0] \equiv 0$
in (\ref{VolterraSeries}) without any loss of generality).  This
use of Volterra series proves to be too restrictive \cite{DY03-2}.
So let us use a more `dynamic' form of the Volterra series
concept.

Instead of expanding the actions in Volterra series like
(\ref{VolterraSeries}), we will now expand the Lagrangians in
Volterra series like \cite{Rugh02}
\begin{eqnarray}\label{LVS-1}
\mathcal{L}[q;\tau] & = & \mysum \frac{1}{n!} K_{n}(\tau)\star
                q^{n}
\end{eqnarray}
where the $\star$ notation is as in (\ref{Knstar}) and the
$K_{n}(\tau)$ will now be of the form
\begin{eqnarray}\label{Kn-1}
K_{n}(\tau) & = & K_{n}(\tau,\tau_{1},\ldots,\tau_{n})    .
\end{eqnarray}
(We assume that $K_{0}(\tau) \equiv 0$ in (\ref{LVS-1}).) We call
a kernel $K_{n}(\tau,\tau_{1}, \ldots, \tau_{n})$
\textit{symmetric} if it is symmetric under any interchange of the
$\tau_{i}, i=1, \ldots, n$.  A kernel is called
\textit{stationary} if there exists a kernel $\kappa_{n}(\xi_{1},
\ldots, \xi_{n})$ such that
\begin{eqnarray}\label{stationarydef}
\kappa_{n}(\tau- \tau_{1}, \ldots, \tau- \tau_{n}) & = &
        K_{n}(\tau,\tau_{1},\ldots,\tau_{n})
\end{eqnarray}
for all $\tau,\tau_{1},\ldots, \tau_{n}$.

The $\mathcal{L}[q,\tau]$ will generally be $\tau$-dependent
distributions that also depend on the function $q$.  So, in a
sense, the Lagrangians can also be viewed as $\tau$-dependent
functionals.  Hence the notation $\mathcal{L}[q;\tau]$ used in
(\ref{LVS-1}). This is also why we allow the extra $\tau$ to
appear in (\ref{Kn-1}) versus (\ref{Ks}).

From (\ref{LVS-1}) we will form the actions
\begin{eqnarray}\label{VVS-1}
\mathcal{V}[q;t] & = & \int_{a}^{t} \mathcal{L}[q;\tau]\ d\tau   .
\end{eqnarray}
The actions given by (\ref{VVS-1}) are $t$-dependent Volterra
series.  So they can be thought of as distributions or
$t$-dependent functionals as the Lagrangians were.  It is also
useful to view the actions as the anti-derivatives of the
Lagrangians.

Our entire problem can now be succinctly stated:  `Find the
correct kernels $K_{n}(\tau,\tau_{1},\ldots,\tau_{n})$ to use in
(\ref{LVS-1}).'  The Volterra series framework helps clarify the
problem of deriving a single causal equation of motion.  It
reduces the whole question to that of finding a correct family of
kernels for our Volterra series.  Our attention now turns to this
problem.

\section{Causal fractional Lagrangian mechanics}\label{SEC-Causal}

Systems and control theorists tend to be obsessed with the
question of causality.  So it seems somewhat natural that the
theory and formalism of causal Volterra series reached its highest
development due to the work of (mathematically oriented)
electrical engineers. This is probably best expressed by the book
written by Rugh \cite{Rugh02}.  We are now going to apply this
previous research to analytic mechanics.  We will see that
enforcing causality in the Lagrangians will result in causal
equations of motion.  By now we should see that the key to
deriving purely causal equations of motion is to find the correct
kernels for our Volterra series.  What we will do in this section
is introduce a new family of kernels to be used in our Volterra
series.  We then examine these kernels to model the
nonconservative harmonic oscillator.  Next, a driving force is
added on.  Finally, higher ordered potentials are dealt with.

First, let us define the symmetric and stationary kernels
\begin{eqnarray}
\Rk{\lambda}{n} & := & \frac{C_{n-1}}{\Gamma (\lambda + n -2)}
        \left( \sqrt{ (\tau-\tau_{1})^{2} + \ldots +
        (\tau-\tau_{n})^{2}}\ \right)^{\lambda-1},    \label{Rdef}
\end{eqnarray}
for $n \geq 2$, and
\begin{eqnarray}
\rk{\lambda}{n} & := & \frac{C_{n}}{\Gamma (\lambda + n -1)}
        \left( \sqrt{ (\tau-\tau_{1})^{2} + \ldots +
        (\tau-\tau_{n})^{2}}\ \right)^{\lambda-1},    \label{rdef}
\end{eqnarray}
for $n \geq 1$, where, in (\ref{Rdef}) and (\ref{rdef}), $\tau_{i}
\leq \tau$ for $i=1,\ldots,n$ and
\begin{eqnarray}\label{Cdef}
C_{n} & := & \frac{2^{n-1} \Gamma (\frac{n}{2})}{ \pi^{n/2}}.
\end{eqnarray}
(It is interesting to compare the distributions in (\ref{Rdef})
and (\ref{rdef}) with the $r^{\lambda}$ examined in
\cite{GeSh64}.)  The restrictions on the $\tau_{i}$ should be
viewed as part of the definitions of the $\Rk{\lambda}{n}$ and the
$\rk{\lambda}{n}$. Requiring that $\tau_{i} \leq \tau$ is what
will give us our causality.  At first sight the distributions in
(\ref{Rdef}) and (\ref{rdef}) appear to have nothing to do with
the fractional derivative operators $\Phi^{\pm}_{\alpha}(\tau)$
presented in section \ref{SEC-FracDer}.  This will prove to not be
the case presently.

The easiest case to deal with is the nonconservative harmonic
oscillator, which we examine now.  Consider the Lagrangian
\begin{eqnarray}\label{LHO}
\mathcal{L}_{HO}[q;\tau] & = & \frac{1}{2} \left[ m \Rk{-2}{2} +
        mC \Rk{-\gamma}{2} + m\omega^{2} \Rk{0}{2}\right] \star
        q^{2}
\end{eqnarray}
where $0 < \gamma < 2$ and $C$ is a constant.  Remember that the
Lagrangian is a $\tau$-dependent Volterra series that will be
treated as a distribution.  From (\ref{LHO}) we form the action
\begin{eqnarray}\label{VHO}
\mathcal{V}_{HO}[q;t] & := & \int_{a}^{t}
            \mathcal{L}_{HO}[q;\tau]\
                            d\tau       .
\end{eqnarray}
In (\ref{VHO}) we take $q(\tau_{i}) \equiv 0$ for $t_{i} < a$.
This implies that $\mathcal{L}[q;\tau] \equiv 0$ for $\tau < a$ in
(\ref{LHO}) since the Lagrangian in (\ref{LHO}) is causal.  Also,
the action in (\ref{VHO}) is a $t$-dependent Volterra series.
Remember that we think of $\mathcal{V}[q;t]$ in (\ref{VHO}) as
being the anti-derivative of $\mathcal{L}[q;\tau]$ in (\ref{LHO}).

In (\ref{LHO}) and (\ref{VHO}) we should think of $\tau$ as being
`now'.  When we derive our equations of motion shortly, we will
vary the $q(\tau_{i})$ in (\ref{VHO}).  When we do this, we always
consider the perturbations of $q(\tau_{i})$ as happening `now',
i.e.,
\begin{eqnarray}\label{qPert}
q(\tau_{i}) & \longrightarrow & q(\tau_{i}) + h \delta(\tau -
            \tau_{i})
\end{eqnarray}
where $h$ is infinitesimal.  Using (\ref{qPert}) in (\ref{VHO})
gives us
\begin{eqnarray}
\frac{\delta \mathcal{V}[q;t]}{\delta q(\tau)} & = & \int_{a}^{t}
    \left[m \rk{-2}{1} + mC \rk{-\gamma}{1} + m\omega^{2}
    \rk{0}{1}\right] \star q\ d\tau \nonumber   \\
    & =& \int_{a}^{t} \left\{ m \LFDt{a}{2}{q} + mC
    \LFDt{a}{\gamma}{q} + m\omega^{2} \LFDt{a}{0}{q} \right\}\
    d\tau    \label{VHOpert}   .
\end{eqnarray}
We will require that (\ref{VHOpert}) is the zero distribution for
$t>a$.  This means that
\begin{eqnarray}\label{HOeom}
m \LFDt{a}{2}{q} + mC
    \LFDt{a}{\gamma}{q} + m\omega^{2} \LFDt{a}{0}{q} & = & 0
\end{eqnarray}
for $\tau>a$.  Equation (\ref{HOeom}) is the nonconservative
harmonic oscillator's equation of motion.  The above can also be
extended to $\Rk{-\lambda}{2}$ where $\lambda > 2$.  So,
derivatives of arbitrary order of $q(\tau)$ can be included in
(\ref{LHO}).

So how did the restriction $\tau_{i} \leq \tau$ result in the
causality in (\ref{HOeom})?  Consider the quantity
\begin{eqnarray}\label{Sex}
\mathcal{S}[q;t] & = & \int \Rk{-\lambda}{2} q(\tau_{1})
        \delta(\tau-\tau_{2})\ d\tau_{2} d\tau_{1} d\tau
        \nonumber           \\
        & = & \int \frac{1}{\Gamma (-\lambda)} \left( \sqrt{(\tau -
        \tau_{1})^{2} + (\tau - \tau_{2})^{2}}\ \right)^{-\lambda-1}
        q(\tau_{1}) \delta(\tau-\tau_{2})\ d\tau_{2} d\tau_{1} d\tau
        \nonumber           \\
        & = & \int \frac{1}{\Gamma (-\lambda)} \left( \sqrt{(\tau -
        \tau_{1})^{2}}\ \right)^{-\lambda-1} q(\tau_{1})\ d\tau_{1} d\tau
        .
\end{eqnarray}
Since $\tau_{1} \leq \tau$, we can rewrite (\ref{Sex}) as
\begin{eqnarray}\label{Sex2}
\mathcal{S}[q;t] & = & \int_{a}^{t} \int_{a}^{\tau} \frac{(\tau -
    \tau_{1})^{-\lambda-1}}{\Gamma (-\lambda)}\ q(\tau_{1})\ d\tau_{1}
    d\tau       \nonumber       \\
    & = & \int_{a}^{t} \PHIp{-\lambda}{\tau} \ast q(\tau)\ d\tau
\end{eqnarray}
where $q(\tau) \equiv 0$ for $\tau < a$.  Hence, the restrictions
on the $\tau_{i}$ allows us to have the $\LFDt{a}{\lambda}{q}$
arise from the use of the $\Rk{-\lambda}{2}$ in the Lagrangians.

Driving forces need to be handled somewhat differently than the
potential or kinetic energy terms.  This is perhaps not surprising
since a driving term is, in a sense, outside the `universe' we are
modelling.  Let us consider the $q(\tau)$ as our system that moves
through some environment described by the kernels in our Volterra
series.  So the Volterra series includes both our system and
environment.  The driving term is neither of these.  We could of
course expand our view of what the environment or system are to
include those mechanisms that give rise to the driving force.
This, however, may overly complicate matters.  Instead, we will
proceed as follows for, e.g., the driven nonconservative harmonic
oscillator.  Let our new Lagrangians be given by
\begin{eqnarray}\label{LdHO}
\mathcal{L}^{'}[q;\tau] & = & \mathcal{L}_{HO}[q;\tau] -
    \int_{a}^{\tau} f(\tau_{1}) q(\tau_{1})\ d\tau_{1}   .
\end{eqnarray}
Things proceed as before.  So our action is given by
\begin{eqnarray}\label{VdHO}
\mathcal{V}^{'}[q;t] & = & \mathcal{V}_{HO}[q;t] - \int_{a}^{t}
    \int_{a}^{\tau} f(\tau_{1}) q(\tau_{1})\ d\tau_{1} d\tau .
\end{eqnarray}
Perturbing $q(\tau_{i})$ by $h\delta(\tau-\tau_{i})$ results in
\begin{eqnarray}\label{VdHOpert}
\frac{\delta \mathcal{V}^{'}[q;t]}{\delta q(\tau)} & = &
    \frac{\delta \mathcal{V}[q;t]}{\delta q(\tau)} - \int_{a}^{t}
    f(\tau)\ d\tau   .
\end{eqnarray}
Requiring (\ref{VdHOpert}) to be the zero distribution for $t>a$
gives us
\begin{eqnarray}\label{dHOeom}
m \LFDt{a}{2}{q} + mC
    \LFDt{a}{\gamma}{q} + m\omega^{2} \LFDt{a}{0}{q} & = & f(\tau)
\end{eqnarray}
for $\tau>a$.  So driving forces are easily included in our
equations of motion via the term
\begin{eqnarray}\label{Drivingterm}
-\int_{a}^{\tau} f(\tau_{1}) q(\tau_{1})\  d\tau_{1}
\end{eqnarray}
in our Lagrangians.

Now we turn our attention to higher ordered potentials.  Consider
a term like
\begin{eqnarray}\label{LHOpot}
\mathcal{L}_{n}[q;\tau] & = & \frac{1}{n}\ \Rk{\lambda}{n} \star
        q^{n}
\end{eqnarray}
in our Lagrangians, where $n \geq 2$.  The term in the action
resulting from (\ref{LHOpot}) is given by
\begin{eqnarray}\label{VHOpot}
\mathcal{V}_{n}[q;t] & = & \frac{1}{n}\ \int_{a}^{t}
    \Rk{\lambda}{n} \star q^{n}\ d\tau       .
\end{eqnarray}
Perturbing $q(\tau_{i})$ as before gives us
\begin{eqnarray}
\frac{\delta\mathcal{V}_{n}[q;t]}{\delta q(\tau)} & = &
    \int_{a}^{t} \rk{\lambda}{n-1} \star q^{n-1}\ d\tau \label{VHOpert1}      \\
    & = & \int_{a}^{t} \int_{0}^{\tau-a} \cdots \int_{0}^{\tau-a}
    \frac{C_{n-1}}{\Gamma (\lambda +n -2)}\ \left(\sqrt{
    \tau_{1}^{2} + \cdots + \tau_{n-1}^{2}}\ \right)^{\lambda-1}
     \nonumber                 \\
    & & \times\ q(\tau-\tau_{1}) \cdots q(\tau-\tau_{n-1})\ d\tau_{n-1}
    d\tau_{1} d\tau          \label{VHOpert2}
\end{eqnarray}
where (\ref{VHOpert2}) follows from (\ref{VHOpert1}) by the change
of variables $\tau_{i} \longrightarrow \tau-\tau_{i}$.  Now we
switch to spherical coordinates so that (\ref{VHOpert2}) becomes
\begin{eqnarray}\label{Vnspher}
\frac{\delta\mathcal{V}_{n}[q;t]}{\delta q(\tau)} & = &
    \int_{a}^{t} \int_{0}^{\infty} \frac{C_{n-1}r^{\lambda+n-3}}{\Gamma(
    \lambda+n -2)}\ \frac{Q(r,\tau)}{C_{n-1}}\ dr d\tau \nonumber
    \\
    & = & \int_{a}^{t} \int_{0}^{\infty} \frac{r^{\lambda+n-3}}{\Gamma(
    \lambda+n -2)}\ Q(r,\tau)\ dr d\tau
\end{eqnarray}
where $Q(r,\tau)$ is that part of the integral in (\ref{VHOpert2})
that depends on $r$ and $\tau$ after integrating over the angles
$\omega_{i}, i=1,\ldots, n-1$.  (Note that the restrictions
$\tau_{i} \leq \tau$ become restrictions on the $\omega_{i}$ when
we switch to spherical coordinates.  Also, the exact form of
$Q(r,\tau)$ is unimportant for our purposes, as we now show.)
Rewrite (\ref{Vnspher}) as
\begin{eqnarray}\label{Vnsper2}
\frac{\delta\mathcal{V}_{n}[q;t]}{\delta q(\tau)} & = &
    \int_{a}^{t}\int_{0}^{\infty} \PHIp{\lambda+n-2}{r} Q(r,\tau)\ dr d\tau  .
\end{eqnarray}
For the potential energy terms we take $\lambda = 2-n$.  Then
(\ref{Vnsper2}) becomes
\begin{eqnarray}\label{Vnspher3}
\frac{\delta\mathcal{V}_{n}[q;t]}{\delta q(\tau)} & = &
    \int_{a}^{t} Q(0,t)\ d\tau  \nonumber       \\
    & = & \int_{a}^{t} q^{n-1}(\tau)\ d\tau .
\end{eqnarray}
From (\ref{Vnspher3}) it follows that (\ref{LHOpot}), with
$\lambda=2-n$, can be rewritten as
\begin{eqnarray}\label{LHOpot2}
\mathcal{L}_{n}[q;\tau] & = & \frac{1}{n}\ \delta\left(\sqrt{
        (\tau-\tau_{1})^{2} + \cdots + (\tau-\tau_{n})^{2}}\ \right) \star
        q(\tau_{1}) \cdots q(\tau_{n})
\end{eqnarray}
so that (\ref{VHOpot}) becomes
\begin{eqnarray}\label{VHOpot2}
\mathcal{V}_{n}[q;t] & = & \frac{1}{n}\ \int_{a}^{t}
        \delta\left(\sqrt{
        (\tau-\tau_{1})^{2} + \cdots + (\tau-\tau_{n})^{2}}\ \right) \star
        \\
        & & q(\tau_{1}) \cdots q(\tau_{n})\ d\tau \nonumber  .
\end{eqnarray}
All our potential energy terms in the Lagrangian will be as in
(\ref{LHOpot2}).

So now we can include driving forces, higher ordered potentials
and derivatives of arbitrary order in our Lagrangians.  We are
also able to derive purely causal equations of motion by using the
kernels in (\ref{Rdef}) in our Lagrangians.  A general Lagrangian
will be given by
\begin{eqnarray}\label{GenL}
\mathcal{L}[q;\tau] & = & \int_{a}^{\tau} f(\tau_{1}) q(\tau_{1})\
    d\tau_{1} + \frac{1}{2}\ \sum_{j=1}^{k} c_{j}
    \Rk{-\alpha_{j}}{2} \star q(\tau_{1}) q(\tau_{2})  + \\
    & & \sum_{m=2}^{\infty} \frac{c_{m}}{m}\ \delta\left(\sqrt{
        (\tau-\tau_{1})^{2} + \cdots + (\tau-\tau_{m})^{2}}\ \right) \star
        q(\tau_{1}) \cdots q(\tau_{m})      \nonumber
\end{eqnarray}
where the $c_{j}$ and $c_{m}$ are constants and $0<\alpha_{1} <
\ldots < \alpha_{n}$.  Allowing $\alpha_{j} < 0$ results in a
fractional diffeo-integral equation of motion.

\section{Discussion}\label{SEC-Discussion}

We have presented a rather general method to model nonconservative
systems.  There are two key observations that allowed this to be
possible. The main point was to find the correct form for the
Volterra series expansion of the Lagrangians
$\mathcal{L}[q;\tau]$.  This form is given in (\ref{GenL}).  In
particular, allowing the Volterra series kernels to depend on the
`extra' parameter $\tau$ allowed us to get around Bauer's
corollary \cite{Baue31} and the result of Dreisigmeyer and Young
in \cite{DY03-2}.  Because of this expanded definition of the
allowed kernels, we were able to find the $\Rk{\lambda}{n}$ that
met our requirement for a causal equation of motion.  That is, we
abandoned using the fractional derivative kernels
$\Phi^{\pm}_{\alpha}(\tau)$ in our Lagrangians and actions, as in
\cite{DY03}. This is the second observation that allowed us to
achieve our goal of a single retarded equation of motion.

There are a few areas of future research that would be interesting
to pursue in relation to this paper.  Let us initially list the
more mathematical of these.  First, are the $\Rk{\lambda}{n}$ the
only kernels that allow us to derive the correct equations of
motion? We have not proved that they are in this paper so, there
may be other, more appropriate kernels to use.  Also, it would be
nice to have a rigorous exploration of the `generalized
convolution' operation that appears in the $\mathcal{L}[q;\tau]$
and $\mathcal{V}[q;t]$.  While we treated both the Lagrangians and
the actions as distributions, this still needs to be given a firm
mathematical foundation.  Finally, we have only let $\lambda \in
\mathbb{R}$ and $n = 2,3,\ldots$ in the $\Rk{\lambda}{n}$.  In
fact, for $n \neq 2$ only $\lambda = 0$ was allowed.  It would be
interesting to extend this to arbitrary $\lambda \in \mathbb{R}$
and $n \geq 2$. Allowing $\lambda \in \mathbb{R}$ for any $n = 2,
3, \ldots$ is relatively straightforward.  Allowing for fractional
dimensions, e.g., $n \geq 2$ with $n \in \mathbb{R}$, is a more
difficult and interesting question.

For physics, the central problem is probably to develop a
Hamiltonian formulation that corresponds to our Lagrangian one.
Also, it would be interesting to examine what effects special
relativity may have in our formalism.  Finally, the question of
quantization, via, say, the path integral, is also open.  All of
these areas of research may cast light on the question of whether
the $\Rk{\lambda}{n}$ are the correct, or only, kernels to use in
our Lagrangians.

\section{Acknowledgements}\label{SEC-Acknowledgements}
The authors would like to thank the NSF for grant \#9732986.


\bibliographystyle{plain}
\bibliography{FracDer}

\end{document}